# SN 1991T: Reflections of Past Glory[1]

Brian P. Schmidt[2] and Robert P. Kirshner[2]

Harvard-Smithsonian Center for Astrophysics, 60 Garden St., Cambridge, MA 02138

Bruno Leibundgut[2]

European Southern Observatory, Karl-Schwarzschild-Strasse 2, D-85748 Garching bei München, Germany

Lisa A. Wells and Alain C. Porter[3]

Kitt Peak National Observatory[4], P.O. Box 26732, Tucson, AZ 85726-6732

Pilar Ruiz-Lapuente and Peter Challis

Harvard-Smithsonian Center for Astrophysics, 60 Garden St., Cambridge, MA 02138

Alexei V. Filippenko

Department of Astronomy, and Center for Particle Astrophysics, University of California, Berkeley, California 94720

Submitted to *The Astrophysical Journal Letters*

## Abstract

We have obtained photometry and spectra of SN 1991T which extend more than 1000 days past maximum light, by far the longest a SN Ia has been followed. Although SN 1991T exhibited nearly normal photometric behavior in the first 400 days following maximum, by 600 days its decline had slowed, and by 950 days the supernova brightness was consistent with a constant apparent magnitude of $m_B = 21.30$. Spectra near maximum showed minor variations on the SN Ia theme which grew less conspicuous during the exponential decline. At 270 days the nebular spectrum was composed of Fe and Co lines common to SNe Ia. However, by 750 days past maximum light, these lines had shifted in wavelength, and were superimposed on a strong blue continuum. The luminosity of SN 1991T at 950 days is more than $9.0 \times 10^{38} (D/13 \text{ Mpc})^2$ ergs s$^{-1}$ with a rate of decline of less than 0.04 mags/100 days. We show that this emission is likely to be light that was emitted by SN 1991T near maximum light which has reflected from foreground dust, much like the light echos observed around SN 1987A.

---

## 1. Introduction

SN 1991T was discovered in NGC 4527 by a host of observers (Waagen et al. 1991) on 1991 April 13, and reached a maximum of $m_V = 11.5$ mag 16 days later (Phillips et al. 1992; Ford et al. 1993). Soon after its discovery, this object was noted as an unusual member of the SN Ia flock. In a class known for exceptionally uniform spectral and photometric evolution, SN 1991T, before maximum, showed abnormally weak spectral features of intermediate mass elements (Filippenko et al. 1992; Ruiz-Lapuente et al. 1992; Phillips et al. 1992; Jeffery et al. 1993), and an unusually broad maximum in its $B$ and $V$ light curves (Phillips et al. 1992; Ford et al. 1993). Furthermore, this black sheep of the SN Ia family appeared to be more luminous at maximum than standard SNe Ia (Filippenko et al. 1992; Phillips 1993), although other views on the absolute magnitude are tenable (Phillips et al. 1992). However, in the weeks after maximum light, SN 1991T began to lose its unique spectroscopic and photometric features, and soon appeared indistinguishable from other well-observed SNe Ia (Filippenko et al. 1992; Ruiz-Lapuente et al. 1992; Phillips et al. 1992) such as SN 1972E (Kirshner et al. 1973), SN 1981B (Branch et al. 1983), SN 1989B (Barbon et al. 1990; Wells et al. 1994), and SN 1990N (Leibundgut et al. 1991). In this letter, we present photometry and spectra of SN 1991T which extend 1040 days past maximum, the longest time for which a SN Ia has been followed to date. These observations show that about 600 days after maximum SN 1991T stopped the exponential fading characteristic for SNe Ia. Although SN 1991T has once again strayed from its sibling SNe Ia, we believe that this object's peculiar late time behavior has nothing to do with the eccentricities seen in its youth, but instead is due to a light echo formed in intervening dust between the supernova and the Earth.

## 2. Observations

Our optical light curve of SN 1991T was determined from CCD images obtained using the Kitt Peak National Observatory's (KPNO) 2.1 m telescope and the Fred Lawrence Whipple Observatory (FLWO) 1.2 m telescope. A sequence of standard stars (Figure 1 [Plate 1]) near the SN was calibrated on two photometric nights with the FLWO 1.2 m and one photometric night on the 2.1m by observing Landolt (1992) standard stars. Transformation coefficients, for each detector, were derived using the method described by Harris, Fitzgerald, & Reed (1983), and used to determine the $BVR_cI_c$ magnitudes of the sequence stars as given in Table 1. The magnitude of the SN was determined by comparing the magnitude of the SN relative to the sequence stars using the point spread fitting program DOPHOT (Schechter, Mateo, & Saha 1993), correcting for the difference in the color between the standard stars and SN using the color terms determined on the photometric nights. To estimate the error in each measurement, we have used DAOPHOT (Stetson 1987) to place several artificial stars of known brightness onto regions of the galaxy similar to that where the SN lies. The error estimate for each observation is then derived from the mean deviation between the artificial stars' known brightnesses and that output by DOPHOT. The derived photometry and error estimates are presented in Table 2 and Figure 2.

Our spectra (Figure 3) were obtained using the Multiple Mirror Telescope (MMT) red channel spectrograph (Schmidt, Weymann, & Foltz 1989). To obtain our last three spectra it was necessary to position the telescope by offsetting to SN 1991T's position, $\alpha = 12^h 31^m 36.^s 89$, $\delta = +02°56'28''.7$ (B1950.0), from a nearby star (Star 4, Table 1). The spectrograph has 180'' slit, allowing us to subtract the background galaxy by fitting it from the region immediately adjacent to the SN. Although this is an effective method to remove the background contamination, it is prone to errors if the background emission varies significantly over the scale of the spatial resolution. After careful inspection of the two dimensional spectra, we believe that the features seen in these spectra are intrinsic to the SN, and are not the result of contamination from any underlying source. The $(B-V)$ colors synthesized from the spectra, using the filter functions of Bessell (1990) and zero points derived from the Vega spectrum of Hayes, Latham, & Hayes (1975), agrees to 0.1 magnitudes with our photometry, providing further support that the shape of the spectrum is correct. The last three spectra appear identical within the noise, and have been combined to increase the signal to noise ratio. The spectra are presented in Figure 3.

## 3. Discussion of Observations

Few SNe Ia have been well observed at late times; however, SN 1991T's photometric behavior is extraordinary. Although the rate of decline ($\gamma$) between 50 and 300 days past maximum ($\gamma_B$=1.47 mag/100$^d$, $\gamma_V$=1.68 mag/100$^d$) is typical for SNe Ia (Turatto et al. 1990), Figure 1 shows that by 600 days after maximum, SN 1991T had stopped the exponential fading characteristic of SNe Ia. Assuming they had the same luminosity at maximum, SN 1991T was intrinsically more than two magnitudes brighter than SN 1972E (Kirshner & Oke 1975) and SN 1990N (Phillips et al. 1994) at 600 days. If SN 1991T was intrinsically brighter than SN 1972E at maximum light as suggested by Filippenko et al. (1992) and Phillips (1993), then its excess luminosity at late times would be even more pronounced. Over our last three photometric observations, between 750 and 950 days after maximum light, SN 1991T's rate of decline was consistent with zero, $\gamma_B = 0.01 \pm 0.03$ mag/$100^d$. The total luminosity of the SN on this late time plateau is substantial, $9 \times 10^{38} (D/13 \text{ Mpc})^2 \left(10^{0.4 A_V}\right)$ ergs s$^{-1}$ being emitted in the optical ($BVRI$) alone.

At 270 days past maximum, SN 1991T's spectrum was similar to other SNe Ia at this epoch (Ruiz-Lapuente & Filippenko 1993). Correcting for the redshift of NGC 4527, lines primarily made up of [FeIII] multiplets are observed at $\lambda_{obs} = 3950$ Å, $\lambda_{obs} = 4690$ Å, $\lambda_{obs} = 4980$ Å, and $\lambda_{obs} = 5240$ Å; multiplets of [FeII] are contributing at $\lambda_{obs} = 4355$ Å, $\lambda_{obs} = 4980$ Å, $\lambda_{obs} = 5240$ Å, and $\lambda_{obs} = 7200$ Å; and lines from [CoIII] are apparent at $\lambda_{obs} = 5880$ Å, and $\lambda_{obs} = 6540$ Å. The width of these lines indicate an expansion velocity of the Fe peak ejecta of $v > 12000$ km s$^{-1}$, which is approximately 20% faster than either SN 1972E (Kirshner & Oke 1975) or SN 1990N (Phillips et al. 1994).

More than a year and a half later, 752 days past maximum, the SN's spectrum was different. SN 1991T appears to be emitting a larger fraction of its flux blueward of 4500 Å, and the feature attributed to several [Fe III] lines, (at $\lambda_{obs} = 4690$ Å at 270 days) is narrower, and substantially blue shifted to $\lambda_{obs} = 4600$ Å. Emission features are also apparent at $\lambda_{obs} = 4920$ Å, and $\lambda_{obs} = 5260$ Å. It is not proven, however, that SN 1991T is different from other SNe Ia spectra at this age; SN 1972E is the only other SN Ia with a spectrum at such an age (Kirshner & Oke 1975), and there is very little information in this observation. Between 752 and 1040 days past maximum, no further evolution in SN 1991T's spectrum was detected.

## 4. Understanding SN 1991T's Late Time Luminosity

There are several possible explanations for SN 1991T's large late time luminosity. Before embarking on explanations for this observation, a skeptical reader might wonder if we are seeing the SN at all. We can be confident that the object we are seeing is indeed SN 1991T (and not some other source) for several reasons. The observed object is a point source (FWHM < 1.1″) located within 0.2″ of the position of SN 1991T. The absolute magnitude of the object, $M_B = -9.3 - 5\log(D/13 \text{ Mpc})$, is equivalent to 12 O5 Ia stars (Allen 1973), and could be a luminous OB association; however, no object is present at this position on a photographic plate taken a decade before the SN explosion (Figure 1 [Plate 1]; Sandage, Binggeli, & Tammann 1984), even though objects as faint as SN 1991T are clearly visible. Furthermore, the object's color $[(B - V) = 0.0, (V - R) = -0.5]$, shows a large $R$ deficiency which is not consistent with the thermal spectrum of hot stars or the H$\alpha$ rich spectrum of an H II region, but is consistent with SNe Ia colors; we also do not see the narrow emission lines associated with H II regions and OB associations. Most convincingly, the strongest spectral features observed in the object's spectrum after 700 days are broad lines like those of a supernova.

The physics of burning material to nuclear statistical equilibrium is well understood, and the resulting isotopic ratios are reasonably insensitive to the expected variations in SNe of neutron fraction, density, and temperature (Woosley 1986). Therefore, although an unusual abundance of $^{44}$Ti, for example, is difficult to rule out observationally, it is implausible that SN 1991T over-produced this isotope by an order of magnitude (more than 0.02 M$_\odot$ would be required). Other isotopes, such as $^{57}$Co, can be ruled out by the extremely slow rate of decline. Strong circumstellar interaction powering the light curve is also unlikely because such a mechanism would probably produce strong lines of hydrogen (among others), which are not observed. Fransson, Houck, & Kosma (1994) have run time-dependent models of SNe Ia through late times. Their calculations show that models of SNe Ia such as DD4 (Woosley 1991) begin to undergo ionization freeze out (cf. Fransson & Kosma 1993 for SN 1987A) at approximately 1000 days, several hundred days later than necessary to explain SN 1991T's late time luminosity. A more general problem with these models, however, is encountered between 450 to 700 days after explosion, when they undergo an IR-catastrophe as described by Axelrod (1980), causing the optical band light curves to fall sharply from their constant rate of exponential decline. This scenario, while difficult to test with SN 1991T, is in disagreement with more typical SNe Ia such as SN 1972E and SN 1990N.

We believe that SN 1991T's late time luminosity can be attributed to a light echo. The possibility of detecting light echos — the reflection of a SN's light from dust in the surrounding ISM — has long been discussed (Zwicky 1940; van den Bergh 1975; Chevalier 1986; Schaefer 1987), but it was not until SN 1987A (Suntzeff et al. 1988; Crotts 1988; Gouiffes et al. 1988) that this phenomenon was unambiguously seen in association with a supernova.[5] The total integrated magnitude of SN 1987A's light echo was $m_V = 13.1$ and remained essentially unchanged over the period between 1988 and 1991 (Crotts, personal communication). This ratio of the SN's maximum light brightness and the light echo brightness (∼9 mag) and the constant flux is consistent with what we see in the case of SN 1991T. Furthermore, we know that there was a significant amount of dust in front of SN 1991T from interstellar absorption lines seen at the recession velocity of the host galaxy (Meyer & Roth 1991).

The light echo hypothesis makes a strong prediction: the spectrum observed in reflected light is expected to consist of light emitted at early times. We should see the intensity-weighted, time-integrated spectrum of SN 1991T modified by preferential scattering of blue light by the dust. We have computed the time-integrated spectrum of SN 1991T for the 75 days following discovery by weighting each spectrum of SN 1991T with its $V$ magnitude (Phillips et al. 1992), and interpolating between spectra to give uniform time coverage. We have parameterized the effects of scattering as a $S_\lambda \propto \lambda^{-\alpha}$ scattering law (Suntzeff, et al. 1988). The results are compared compared to SN 1991T's late time spectrum in Figure 4. The match is very good using $\alpha = 2$, with the major features in the spectrum all represented, and at the correct wavelength. Because we could only integrate the spectrum of SN 1991T for the 75days following discovery (for lack of observations) it is not surprising that some of the lines (e.g., that at 4600 Å) do not appear sufficiently strong in the integrated spectrum. It is worth noting that the time-integrated spectrum of SN 1991T is different from its maximum light spectrum, and simply comparing the echo spectrum with the maximum light spectrum would be misleading.

At late times, SN 1991T is relatively bright and constant in luminosity. These observations lead us to believe that the light echo is being formed in a cloud of dust in NGC 4527 which is tens or hundreds of light years in front of SN 1991T, and not with dust associated with the supernova. If the reflecting dust were close to the supernova, the scattering angle for the light would be large (very inefficient compared to forward scattering) and would change significantly over a year (thereby changing the brightness of the echo). If the light echo is being formed in a foreground dust cloud, it probably has an apparent size greater than 0.1″, and should be resolvable using the Hubble Space Telescope. Such observations would determine the position of the intervening cloud. Unfortunately, it is unlikely that a direct, geometric distance determination as discussed by Sparks (1994) will be possible for this object. His method exploits the fact that the polarization of scattered light has a sharp maximum at scattering angles of 90°. He demonstrates with this geometry that, assuming there is sufficient dust located in the appropriate position in space, the light echo in polarized light will have a radius $r = ct$. Unfortunately, the dust producing SN 1991T's light echo is most likely located well in front of the supernova, and therefore the dust will not be located close enough to the SN to produce the ring of polarized light.

---

[5] Schaeffer (1987) claimed to see evidence for a light echo around SN 1986G from a slowing in its light curve by 400 days. However, Phillips (personal communication) has re-analyzed these data and concludes that SN 1986G continued to fall at its previous rate, and did not slow in its decline.

We believe this is, after SN 1987A, the second conclusive detection of a supernova light echo. However, there is no reason to think that SN 1991T's light echo has anything to do with its early idiosyncrasies. Careful inspection of other bright supernovae (of all types) which have some extinction from their host galaxy (SN 1993J in M81 and SN 1979C in M100 come to mind) might yield similar findings to those presented here.

This work is dedicated to the memory of Alain Porter, whose hard work and determination will not be forgotten, whose contributions we valued, and whose companionship in scientific adventures we will miss. We would also like to thank Bruno Bingelli for providing us with the pre-explosion image of NGC 4527, as well as Mark Phillips for providing his early-time SN 1991T data and for his helpful comments. We are grateful for valuable discussions with Philip Pinto and Peter Höflich. This work is supported in part by NSF grants AST 92-18475 and AST 91-15174, as well as NASA grants NAG 5-841 and NGT-51002.

# References

Allen, C. W. 1973, Astrophysical Quantities, 3rd ed., (London: Athlone Press)

Barbon, R., Benetti, S., Rosino, L., Cappellaro, E., & Turatto, M. 1990, A&A, 237, 79

Bessell, M. S. 1990, PASP, 102, 1181

Branch, D., Lacy, C., McCall, M.L., Uomoto, A., Wheeler, C, Wills, B. J., & Sutherland, P.G, 1983, ApJ, 270, 123

Chevalier, R. A. 1986, ApJ, 308, 225

Crotts, A. P. S. 1988, ApJ, 333, L51

Crotts, A. P. S., Kunkel, W. E., & McCarthy, P. J. 1989, ApJ, 347, L61

Filippenko, A. V., et al. 1992, ApJ, 384, L15

Ford, C. H., Herbst, W., Richmond, M. W., Baker, M. L., Filippenko, A. V., Treffers, R. R., Paik, Y., & Benson, P. J. 1993, AJ, 106, 1101

Fransson, C., & Kozma, C. 1993, ApJ, 408, L25

Fransson, C., Houck, J., & Kozma, C. 1994, IAU Coll. No. 145, "Supernovae and Supernova Remnants, Xian, China" ed. R. McCray, (Cambridge: Cambridge University Press) in press

Harris, W. E., Fitzgerald, M. P., & Reed, B. C. 1981, PASP, 93, 507

Hayes, D. S., Latham, D. W., & Hayes, S. H. 1975, ApJ, 197, 587

Gouiffes, C., Melnick, J., Remy, M., Rosa, M. & Danziger, I. J. 1988, A&A, 198, L9

Kirshner, R. P., & Oke, J. B. 1975, ApJ, 200, 574

Landolt, A. U. 1992, AJ, 104, 340

Leibundgut, B., Kirshner, R. P., Filippenko, A.V., Shields, J.C., Foltz, C.B., Phillips, M.M., & Sonneborn, G. 1991, ApJ, 371, L23

Meyer, D. M., & Roth, K. C. 1991, ApJ, 383, L41

Phillips , M. M. 1993, ApJ, 413, L105

Phillips, M. M., Wells, L. A., Suntzeff, N. B., Hamuy, M., Leibundgut, B., Kirshner, R. P., & Foltz, C. B. 1992, AJ, 103, 1632

Phillips, M. M., et al. 1994, in preparation

Sandage, A., Binggeli, B., & G. A. Tammann 1985, AJ, 90, 395

Ruiz-Lapuente, P. et al. 1992, ApJ, 387, L33

Ruiz-Lapuente, P. & Filippenko, A. V. 1993, in "Origin and Evolution of the Elements", ed. N. Prantzos, E. Vangioni-Flam, & M. Cassé, (Cambridge, University Press) p 318

Schechter, P. L., Mateo, M., & Saha, A. 1993, PASP, 105, 1342

Schmidt, G. D., Weymann, R. J., & Foltz, C. B. 1989, PASP, 101, 713

Schaefer, B. E. 1987, ApJ, 323, L47

Sparks, W. B. 1994, ApJ, 20 Sep. issue

Suntzeff, N. B., Heathcote, S., Weller, W. G., Caldwell, N., & Huchra, J. P. 1988, Nature, 334, 135

Stetson, P. B., 1987, PASP, 99, 191

Turatto, M., Cappellaro, E., Barbon, R., Della Valle, M., Ortolani, S., & Rosino, L. 1990, AJ, 100, 771

van den Bergh, S. 1975, Ap.Space.Sci., 38, 447

Waagen, E. et al. 1991, IAU Circ. #5239

Wells, L. A. et al. 1994, AJ, in press

Woosley, S. E. 1986, in "Nucleosynthesis and Chemical Evolution", ed B. Hauck & A. Maeder, (Geneva Observatory, CH-1290, Sauverny, Switzerland)

Woosley, S. E. 1991, Gamma-Ray Line Astrophysics, ed. P. Durouchoux & N. Prantzos, (New York: American Institute of Physics), 270

Zwicky, F. 1940, Rev. of Mod. Physics, 12, 66

TABLE 1. Comparison Star Magnitudes

| Star | $\alpha$ | $\delta$ | B | V | R | I |
|------|----------|----------|---|---|---|---|
|      | (B1950.0) |         |   |   |   |   |
| SN   | $12^h31^m36^s.89$ | $+2°56'28''.7$ | ... | ... | ... | ... |
| 1    | $12^h31^m31^s.50$ | $+2°53'58''.8$ | 17.82(04) | 16.34(04) | 15.41(08) | 14.29(08) |
| 2    | $12^h31^m34^s.17$ | $+2°53'48''.6$ | 17.94(03) | 16.78(04) | 16.03(08) | 15.36(08) |
| 3    | $12^h31^m42^s.65$ | $+2°57'11''.9$ | 18.96(04) | 18.52(04) | 18.17(08) | 17.62(08) |
| 4    | $12^h31^m41^s.12$ | $+2°56'43''.3$ | 20.03(05) | 18.63(06) | 17.63(09) | 16.60(08) |

Uncertainties in hundredths of magnitudes are listed in parentheses

TABLE 2. Late-Time Photometry of SN 1991T

| Date | JD | B | V | R | I | (B-V) | (V-R) | (V-I) | Telescope | Observer |
|---|---|---|---|---|---|---|---|---|---|---|
| 91 Dec 7 | 8597.8 | 17.20(03) | 17.13(03) | 17.65(05) | 17.34(08) | 0.07(04) | -0.61(06) | -0.30(09) | 1.2m | Schmidt |
| 92 Mar 8 | 8688.8 | ... | 18.35(03) | 19.05(05) | 18.43(05) | ... | -0.79(06) | -0.17(06) | 2.1m | Leibundgut |
| 92 Apr 30 | 8742.8 | 19.00(03) | 19.05(03) | 19.64(08) | 18.81(09) | -0.05(04) | -0.68(09) | 0.15(10) | 2.1m | Porter |
| 92 Jun 3 | 8776.8 | 19.43(03) | 19.49(03) | 20.03(08) | 19.08(12) | -0.06(04) | -0.63(09) | 0.32(13) | 2.1m | Porter |
| 93 Jan 26 | 9013.8 | 21.02(05) | 21.01(06) | 21.41(18) | >21.1 | 0.01(08) | -0.49(19) | ... | 2.1m | Wells |
| 93 Mar 21 | 9067.9 | 21.20(08) | ... | ... | ... | ... | ... | ... | 2.1m | Wells |
| 93 May 25 | 9132.8 | 21.28(07) | ... | ... | ... | ... | ... | ... | 2.1m | Wells |
| 93 Dec 10 | 9332.0 | 21.30(15) | 21.43(15) | ... | ... | -0.13(21) | ... | ... | 2.1m | Wells, Schmidt |

Uncertainties in hundredths of magnitudes are listed in parentheses

Figure 1: *B* CCD image of SN 1991T taken with the KPNO 2.1m on 1993 Jan 26. East is up and north is to the right. The locations of the SN and the comparison stars are shown. Precise positions of the SN and comparison stars are given in Table 1. Also shown is a plate of NGC 4527 taken by Sandage et al. (1984) in 1982. The SN position is marked, as well as a star which was of a comparable brightness to the supernova on 1993 Jan 26.

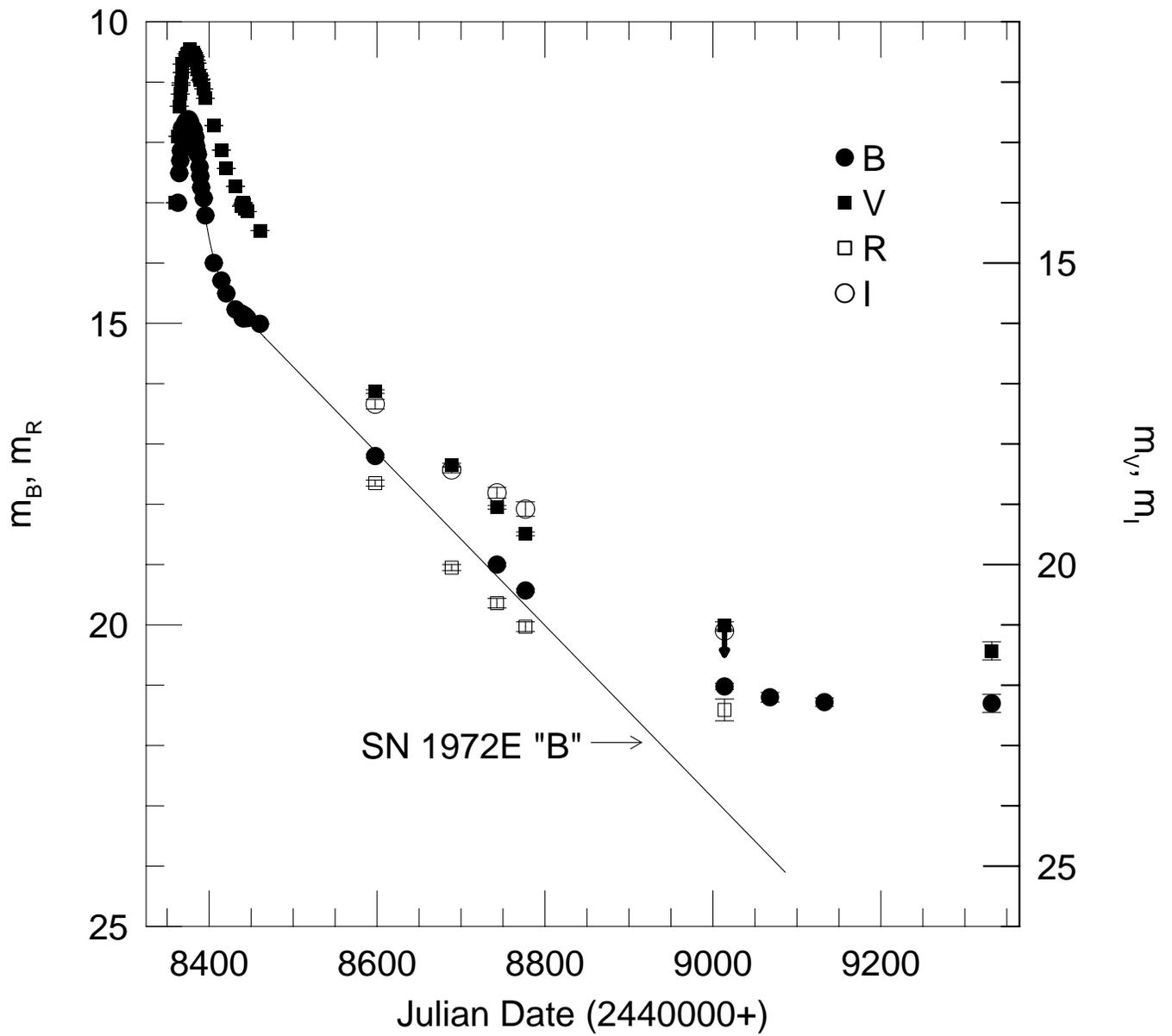

Figure 2: *BVRI* photometry of SN 1991T. Also shown is the early *BV* photometry of SN 1991T (Phillips et al. 1992), and a template of *B* photometry derived from SN 1972E (Kirshner & Oke 1975). The *V* and *I* scales are offset by one magnitude from the *B* and *R* scales.

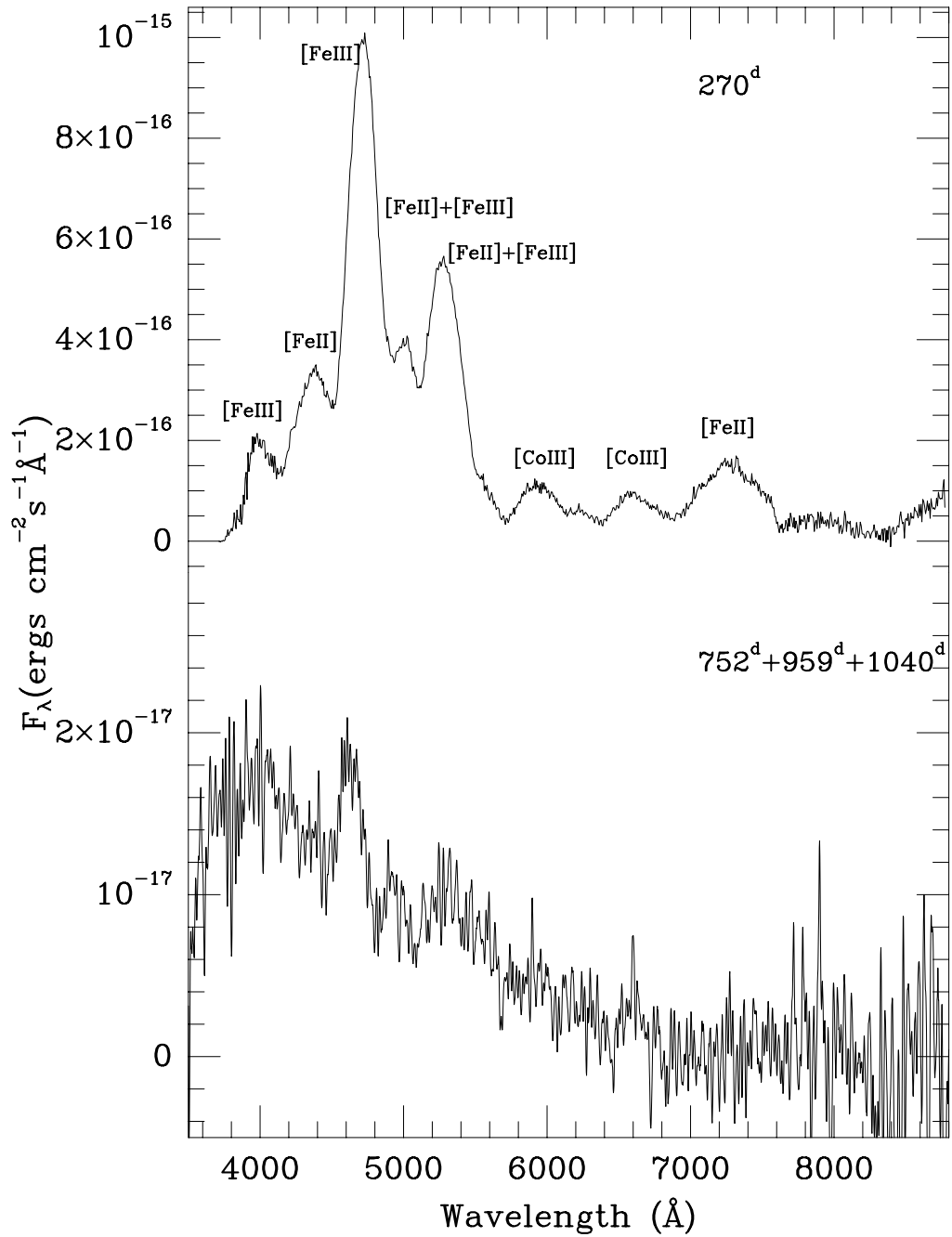

Figure 3: Late time spectral observations of SN 1991T are displayed for 270, 752, 959, and 1040 days after maximum light, corrected for the recession velocity of NGC 4527. The last three spectra appear identical, and have been combined to increase the signal to noise ratio. Suggestions as to the principal contributing ions for each line are included.

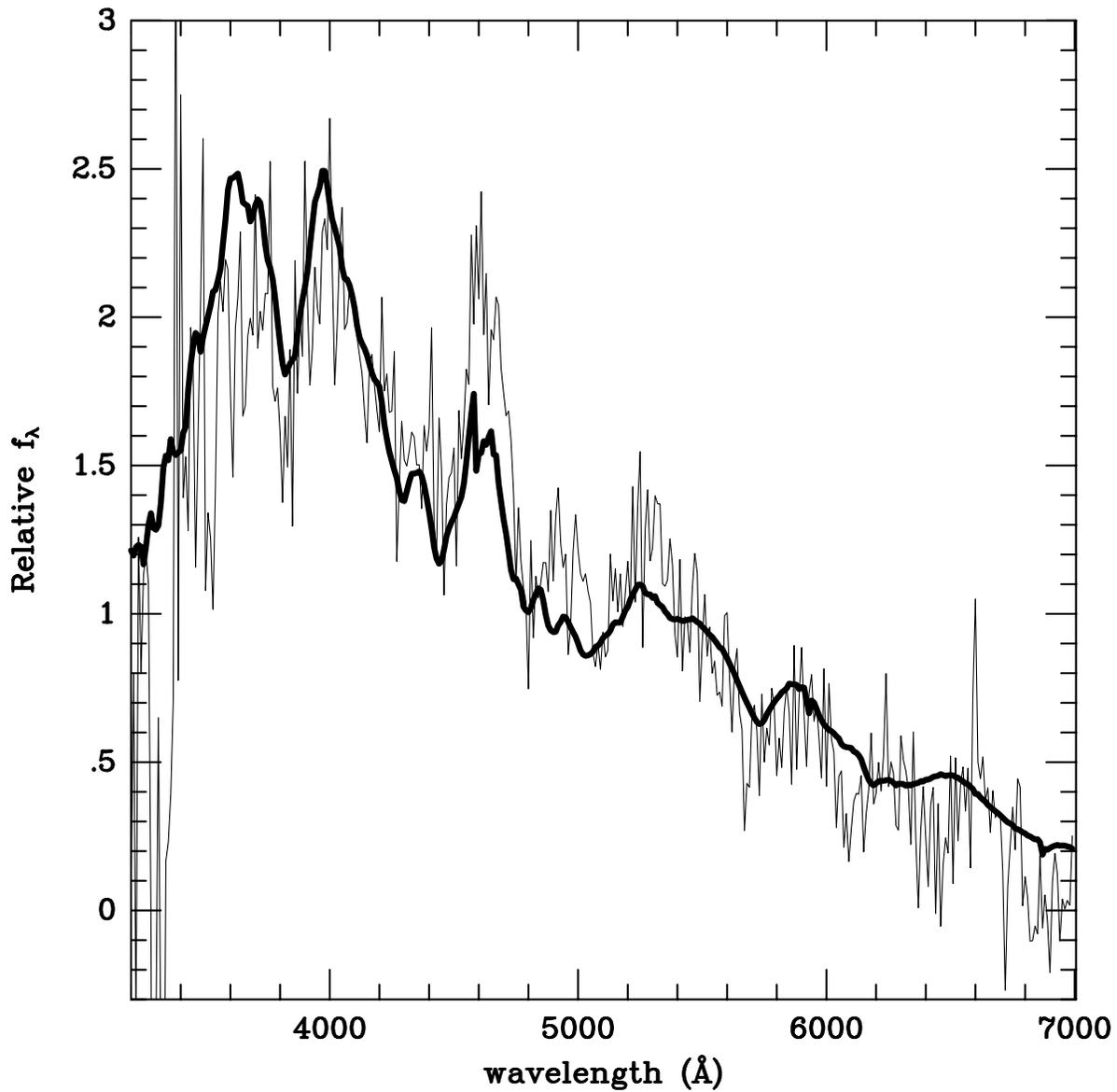

Figure 4: A comparison of the time-integrated, intensity-weighted spectrum with the observed late-time spectrum for SN 1991T. The integrated spectrum has been corrected for the effect of scattering using a $\lambda^{-2}$ law. The good agreement between these two spectra suggests that we are seeing light emitted by SN 1991T around maximum light that has been reflected off intervening dust in the supernova's host galaxy, NGC 4527.